\documentclass[conference]{IEEEtran}
\usepackage[utf8]{inputenc}
\IEEEoverridecommandlockouts
\usepackage{cite}
\usepackage{amsmath,amssymb,amsfonts}
\usepackage{algorithmic}
\usepackage{graphicx}
\usepackage{textcomp}
\usepackage{comment}
\usepackage{xcolor}
\usepackage{amsmath,amssymb,amsfonts}
\usepackage{graphicx}
\usepackage{algorithmic}
\usepackage{graphicx}
\usepackage{textcomp}
\usepackage{lipsum}
\usepackage{amsmath}
\usepackage[numbers]{natbib}
\usepackage{url}
\usepackage[section]{placeins}
\usepackage{authblk}
\usepackage{float}
\usepackage{array}
\usepackage{tabularx}
\usepackage{caption}
\def\BibTeX{{\rm B\kern-.05em{\sc i\kern-.025em b}\kern-.08em
    T\kern-.1667em\lower.7ex\hbox{E}\kern-.125emX}}
  
\begin{document} 
\title{Blockchain Technology in Higher Education Ecosystem: Unraveling the Good, Bad, and Ugly \vspace{-4mm}
}

\begin{comment} 
\author{
\IEEEauthorblockN{Sharaban Tahora}
\IEEEauthorblockA{\textit{dept.  of Information Technology} \\
\textit{Kennesaw State University}\\
USA \\
stahora@students.kennesaw.edu}
\and
\IEEEauthorblockN{Bilash Saha}
\IEEEauthorblockA{\textit{dept.  of Information Technology} \\
\textit{Kennesaw State University}\\
USA \\
bsaha@students.kennesaw.edu}
\and 
\IEEEauthorblockN{Nazmus Sakib}
\IEEEauthorblockA{\textit{dept.  of Information Technology} \\
\textit{Kennesaw State University}\\
USA \\
nsakib1.kennesaw.edu}

\IEEEauthorblockN{Hossain Sahriar}
\IEEEauthorblockA{\textit{dept.  of Information Technology} \\
\textit{Kennesaw State University}\\
USA \\
hshahria@kennesaw.edu}
}
\end{comment}

\author[1]{Sharaban Tahora}
 \author[1]{Bilash Saha}
 \author[1,*]{Nazmus Sakib} 
 \author[1]{Hossain Shahriar} 
\author[2]{Hisham Haddad \vspace{-4mm}}
\affil[1]{Department of Information Technology, Kennesaw State University, Georgia, United States}
\affil[2]{Department of Computer Science, Kennesaw State University, Georgia, United States}
% \affil[]{\{stahora bsaha\}@students.kennesaw.edu}\affil[*]{\{nsakib1 hshahria\}@kennesaw.edu \vspace{-3mm}}

%\author[1]{Authors' Names \vspace}
% \author[1]{Bilash Saha}
% \author[*]{Nazmus Sakib} 
% \author[*]{Hossain Shahriar  \vspace{-3mm}} 

 %\affil[1*]{Authors' Affiliations }
 %\affil[*]{Corresponding Author's Email Address \vspace{-5mm}}
\affil[*]{{nsakib1@kennesaw.edu} \vspace{-3mm}}

\vspace{-6mm}

\maketitle

\begin{abstract}
The higher education management systems first identified and realized the trap of pitting innovation against privacy while first addressing COVID-19 social isolation challenges in 2020. In the age of data sprawl, we observe the situation has been exacerbating since then. Integrating blockchain technology has the potential to address the recent and emerging challenges in the higher education management system. This paper unravels the Good (scopes and benefits), Bad (limitations), and Ugly (challenges and trade-offs) of blockchain technology integration in the higher education management paradigm in the existing landscape. Our study adopts both qualitative and quantitative approaches to explore the experiences of educators, researchers, students, and other stakeholders and fully understand the blockchain's potential and contextual challenges. Our findings will envision an efficient, secure, and transparent higher education management system and help shape the debate (and trade-offs) pertaining to the recent shift in relevant business and management climate and regulatory sentiment.
%Overall, this research aims to contribute to the growing body of knowledge on the use of blockchain in education and to provide insights and recommendations for educators, policymakers, and other stakeholders interested in using blockchain to improve the education system.
\end{abstract}


\section{Introduction}
Blockchain technology is increasingly seen as a disruptive force in the e-learning market, such as areas of digital credentials, student records management, and online payment systems. The e-learning market is expected to reach \$376 billion by 2026, with a growing demand for Massive Open Online Courses (MOOCs) which are expected to be worth \$25.33 billion by 2025 \cite{stats}. MarketsandMarkets estimates that the global blockchain in education market was valued at \$59.7 million in 2019 and is expected to reach \$1,381.9 million by 2023 with a Compound Annual Growth Rate (CAGR) of 84.3\% \cite{stats2}. EdTechXGlobal and HolonIQ found that most universities are exploring the use of blockchain in education management system, with a focus on digital credentialing and student data management \cite{stats3}. Massachusetts Institute of Technology (MIT) established its Blockchain Education Alliance to explore educational applications of blockchain in 2018 \cite{stats4}, and Southern New Hampshire University launched a blockchain-based platform for verifying and sharing student credentials in 2019 \cite{stats5}. The Open University in the UK partnered with blockchain company Learning Machine to launch a pilot program for secure digital credentialing in 2020 \cite{state6}, and the Chinese city of Hangzhou launched a blockchain-based "digital education passport" to store and verify students' academic records and achievements \cite{state7}. The University of Nicosia in Cyprus became the first accredited university to offer a Master's degree in Digital Currency based on blockchain technology in 2021 \cite{state8}, and the University of Bahrain began using blockchain to issue digital degrees the same year \cite{state9}. The blockchain-based platform "Credly" is being used by many educational institutions to issue digital credentials such as certificates, badges, and micro-credentials \cite{state10}. These statistics demonstrate the proliferating implementation of blockchain technology in higher education as a mechanism to amplify the transferability, validity, and credibility of academic achievements and credentials, thereby emphasizing the exigency for additional empirical investigations to explicate its boundaries and streamline its assimilation. 

The higher education management system has been criticized for its inefficiencies and security issues, including an inadequate response to the COVID-19 pandemic, limited resources and outdated infrastructure, limited storage, and privacy concerns. These challenges have led to difficulties in managing and accessing educational data, as well as concerns about data protection and cyber attacks. Blockchain technology can enhance the higher education management system by providing a secure, tamper-proof record of student achievements and reducing fraud \cite{b10}, \cite{b40}. The technology's immutability and smart contracts automate processes, creating an efficient and permanent record without intermediaries \cite{b19}, \cite{b20}. Decentralized architecture eliminates the need for a central authority, creating a secure and tamper-resistant system \cite{b21}, \cite{b13}, \cite{b24}. Popular frameworks for instance the blockchain manifesto framework \cite{b9}, Student-Centered iLearning Blockchain framework (SCi-B) \cite{b58}, and the use of root Merkle methods \cite{b47} can improve learning and activity tracking in higher educational management. Despite recognizing the benefits of blockchain technology, educational institutions face challenges implementing it due to limited expertise and technology limitations\cite{b6}, \cite{b8}.

The aim of this research is to evaluate the potential of blockchain technology in enhancing the efficiency and security of the higher education management system through the use of its decentralized architecture, immutability, and smart contract functionality. In order to achieve this objective, we addressed the following research questions:

\begin{itemize}
    \item RQ1: What are the scopes of blockchain technology in improving the efficiency, security, and transparency of the higher education management system?
    \item RQ2: What are the potential benefits of implementing blockchain technology in the higher education management sector and how can they be measured and evaluated?
    \item RQ3: What are the limitations and challenges of implementing blockchain technology in the higher education management sector and how can they be overcome?
\end{itemize}

To answer these questions, We reviewed 115 articles and conducted a comprehensive examination of the current higher education management system, identifying numerous challenges such as issues with data integrity, access, and student identification due to centralized databases. We explored the potential of blockchain technology as a transformative tool to elevate the higher education management system, utilizing decentralized storage of student data, improving data access and integrity, and verifying student identities and achievements through blockchain-based credentials to combat fraud and promote transparency. Additionally, we identified several future research directions in blockchain technology in higher education management, including technical and legal implications, best practices for implementation, and the optimal technology to achieve desired impact and advancements.

Overall, this article aims to provide a comprehensive overview of the impact of blockchain technology on the higher education management system. It focuses on the opportunities and challenges posed by blockchain, and the implications of its implementation. The paper will examine the different aspects of blockchain and its application in higher education management system and address how it can be used to enhance efficiency and security.

\section{METHODS}

We bring to the forefront the scoping criteria, systematic literature search, and data analysis process utilized in our study to underscore their significance and emphasize these aspects of our research.

\subsection{Scoping criteria}
In the context of the higher education management system, the scoping criteria for implementing blockchain technology refer to the specific parameters and considerations that need to be taken into account when developing a blockchain-based solution for educational purposes. These criteria include the type of data used and stored on the blockchain, security and privacy measures, scalability and performance, regulatory and compliance requirements, and user needs \cite{b4}, \cite{b11}, \cite{b14}. In our research, we considered the challenges in documentation within the higher education management system and investigated the potential of blockchain technology to provide a solution. For instance, when developing a blockchain-based algorithm for secure and transparent record-keeping of student grades and credentials \cite{b14}, we emphasized the importance of considering scoping criteria such as the type of data used and stored on the blockchain, including student names, grades, and other qualifications, as well as information related to the classes and schools they have attended \cite{b55}. Security and privacy were also crucial considerations, and we ensured that the algorithm could protect student data from unauthorized access and share it only with authorized parties, such as educational institutions and potential employers \cite{b57}. Additionally, scalability and performance were taken into account to ensure that the algorithm could handle a large number of transactions and users, making the system fast and efficient \cite{b58}, \cite{b61}. Regulatory and compliance requirements were also considered to ensure that the algorithm met any relevant data privacy regulations \cite{b59}.

\subsection{Systematic Literature Search}
\begin{figure*}[h!]
\centerline{\includegraphics[scale=0.40]{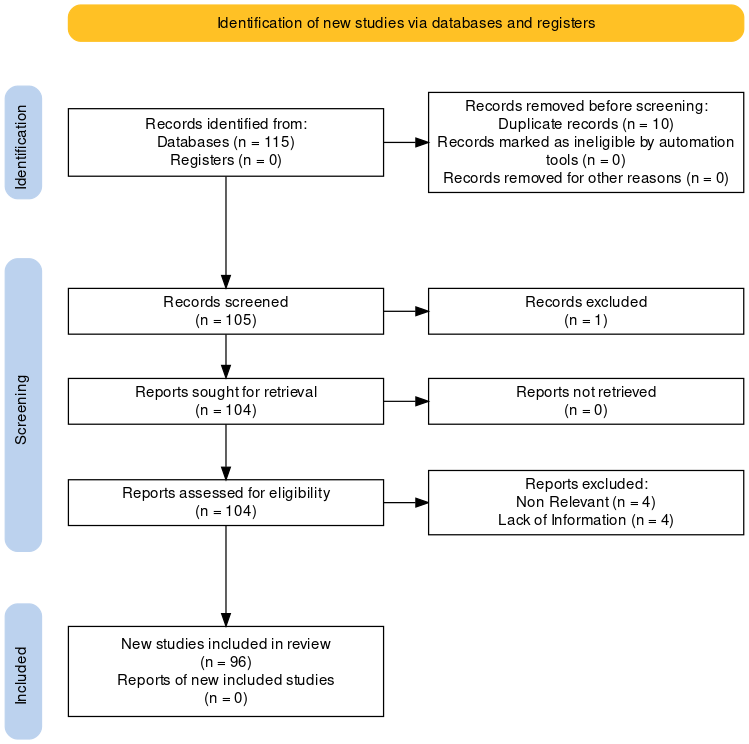}}
\caption{PRISMA flow diagram for study selection in the systematic review}
\label{fig:prisma}
\end{figure*}

A systematic literature review in the realm of blockchain in higher education management system was executed through various techniques. We commenced by identifying pertinent databases, namely JSTOR, Google Scholar, and IEEE Xplore, and then formulated a series of specific search terms that included "blockchain in education", "blockchain and e-learning", "blockchain and education", "blockchain and online learning", "blockchain and degree verification", "Blockchain and credential verification", "Blockchain and education supply chain management", "Blockchain and academic integrity", "Blockchain and educational data management", "Blockchain and smart contract in education", "Blockchain and educational administration", "Blockchain and student data privacy", "Blockchain and e-portfolio", "Blockchain and digital identity management in education", "Blockchain and educational credentialing", "Blockchain in online education", "Blockchain-based learning management systems". The literature search started by searching databases and eliminating irrelevant papers through abstract and title reviews. Key data such as authors, titles, publication date, and findings were then extracted from relevant papers and analyzed to synthesize common themes, trends, and gaps in the literature. The PRISMA flowchart utilized for study selection is shown in Figure \ref{fig:prisma}.

\subsection{Data Analysis Process}
To interpret our findings, we employed a systematic qualitative examination to address our overarching research questions. Specifically, for RQ1 we employed a case study methodology \cite{b63} to investigate specific instances of blockchain-based student information management systems and their implementation in real-world scenarios. For RQ2, we utilized discourse analysis \cite{b64}, \cite{discourse_analysis} to examine the perspectives and views of diverse stakeholders such as educators, students, and employers on the use of blockchain-based credentials. And for RQ3, we adopted an ethnographic approach \cite{b65} to understand the usage of decentralized learning platforms by educators and students and the advantages and drawbacks of this method.

\section{Results}
This section presents our literature review's key findings, including an analysis of the current higher education management system's inadequacies, an exploration of blockchain technology's potential benefits, and an examination of the challenges and strategies for implementing it in the sector.

\subsection{Scoping Blockchain in Higher Education Management System (RQ1)}
The implementation of blockchain technology in the higher education management system has the potential to transform the way information is stored, managed, and transmitted. According to the authors of the paper \cite{b35}, this technology provides a secure, decentralized ledger that is resistant to tampering and hacking, making it an attractive solution for the higher education management sector. Blockchain technology has the potential to bring numerous benefits to the higher education management system, including improved efficiency, security, and transparency. We employed a case study methodology to investigate the use of blockchain-based student information management systems. Table \ref{tab:RQ1} presents an overview of the scope, complexities, benefits, and implementation challenges of this technology in real-world scenarios. These findings allowed us to provide a comprehensive and detailed analysis of the topic, and to better understand the prospects of blockchain technology in the higher education management sector \cite{b24}.  

\begin{figure}[h!]
\centerline{\includegraphics[scale=0.4]{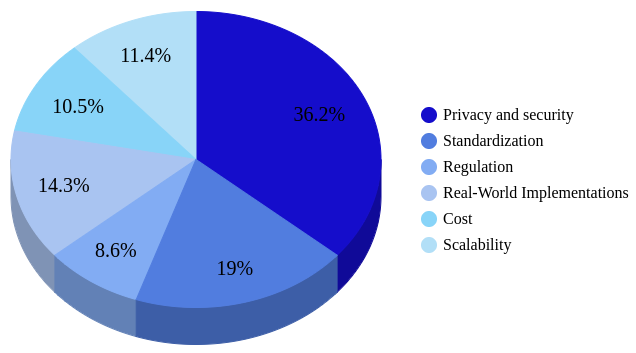}}
\caption{Distribution of Blockchain Implementation in Education Systems}
\label{fig:RQ1}
\end{figure}

Blockchain technology has the potential to address many of the concerns related to privacy and security in the higher education management system. Al Harthy et al. \cite{b27} explained that blockchain technology can provide a secure and transparent way to store data through encryption techniques, which can ensure that personal data remains private and only accessible to authorized parties. The authors in paper \cite{b5} discussed the use of smart contracts to automate various administrative processes in the higher education management systems, such as the issuance of certificates, academic transcripts, and degree verification. Blockchain can also provide a secure and decentralized identity management system \cite{b46} that is tamper-proof and can prevent identity theft. Furthermore, blockchain can eliminate the need for a centralized system by allowing multiple parties to have access to the same data at the same time without a central authority \cite{b52}. This can help to reduce the risk of a single point of failure and increase transparency. Additionally, blockchain technology provides a transparent and immutable record of all transactions, which can address the issue of lack of data regulation \cite{b50}. Figure \ref{fig:RQ1} effectively demonstrates how blockchain can address existing challenges related to privacy and security.

\begin{table*}[ht!]
\footnotesize
\centering
\setlength{\tabcolsep}{0.5em}
{\renewcommand{\arraystretch}{1.2}
 \begin{tabularx}{\textwidth}{X|l|l|l}
\hline
\textbf{Scope} &
  \textbf{\begin{tabular}[c]{@{}l@{}}Challenges and Limitations of \\ Present Higher Education \\ Management System\end{tabular}}  &
  \textbf{Related Sector} &
  \textbf{References} \\ \hline
Privacy and Security &
  \begin{tabular}[c]{@{}l@{}}Data Privacy\\ Identity Management\\ Centralized Systems\\ Lack of Data Regulation\\ Insecure of Data Sharing\end{tabular} &
  \begin{tabular}[c]{@{}l@{}}Access Control \\ Authentication \\ Authorization\\ Data Backup and Recovery \\ Compliance Management\end{tabular} &
  \begin{tabular}[c]{@{}l@{}}\cite{b2},\cite{b4},\cite{b5},\cite{b12},\cite{b16},\cite{b19},\\ \cite{b20},\cite{b21},\cite{b22},\cite{b23},\cite{b24},\cite{b27},\\ \cite{b40},\cite{b41},\cite{b46},\cite{b47},\cite{b50},\cite{b53},\\ \cite{b56},\cite{b59},\cite{b63}
\end{tabular} \\ \hline
Standardization &
  \begin{tabular}[c]{@{}l@{}}Integration Difficulty\\ Harmonization Limitation\\ Availability Barrier\\ Compliance Hurdle\\ Utilization Shortage\end{tabular} &
  \begin{tabular}[c]{@{}l@{}}Learning Content Management System\\ Learning Record Store\\ Learning Resource Metadata\\ Learning Tools Interoperability\end{tabular} &
  \begin{tabular}[c]{@{}l@{}}\cite{b9},\cite{b10},\cite{b12},\cite{b27},\cite{b28}\cite{b54},\cite{HIPPA},\\ \cite{b51},\cite{b54},\cite{b58},\cite{b63}
\end{tabular} \\ \hline
Regulation &
  \begin{tabular}[c]{@{}l@{}}Legal Vagueness\\ Non-compliance\\ Data Disregard\\ Data Mismanagement\\ Inadequate Smart Contract Management\\ Unregulated Cybersecurity\\ Inadequate Auditing.\end{tabular} &
  \begin{tabular}[c]{@{}l@{}}Policy Management\\ Compliance Management\\ Data Governance\\ Risk Management\\ Incident Management\end{tabular} &

  \cite{b22},\cite{b44},\cite{b51},\cite{b59},\cite{b71}\\ \hline
Real-World Implementations &
  \begin{tabular}[c]{@{}l@{}}Lack of Understanding\\ Limited Applications\\ Difficulty in Identifying Suitable Use Cases\\ Limited Research\\ Resistance to Change\\ Limited Interoperability\end{tabular} &
  \begin{tabular}[c]{@{}l@{}}Use Case Identification\\ Learning Analytics\\ Adaptive Learning\\ Personalized Learning \\ Collaborative Learning\end{tabular} &
  \begin{tabular}[c]{@{}l@{}}\cite{b27},\cite{b50},\cite{b51},\cite{b52},\cite{b53},\cite{b56},\\ \cite{b57},\cite{b61}
\end{tabular} \\ \hline
Cost &
  \begin{tabular}[c]{@{}l@{}}Infrastructure Costs\\ Hardware and Software Costs\\ Development Costs\\ Training Costs\\ Maintenance Costs\\ Unclear Return On Investment (ROI)\end{tabular} &
  \begin{tabular}[c]{@{}l@{}}Financial Management\\ Budget Management\\ Cost Benefit Analysis\\ Return On Investment Analysis\end{tabular} &
  \cite{b21},\cite{b27},\cite{b40},\cite{b51},\cite{b55},\cite{b71}
  \\ \hline
Scalability &
  \begin{tabular}[c]{@{}l@{}}Limited throughput\\ High Latency\\ Lack of Data Storage\\ Complexity\\ Energy Consumption\end{tabular} &
  \begin{tabular}[c]{@{}l@{}}Load Balancing\\ Caching\\ Content Delivery Network\\ Data Replication\\ Data partitioning\\ Asynchronous Processing\end{tabular} 
  &
  \cite{b6}, \cite{b13},\cite{b24},\cite{b40}
   \\ \hline
\end{tabularx}%
}
\caption{Scopes of blockchain in the higher education management paradigm}
\label{tab:RQ1}
\vspace{-4mm}
\end{table*}

Blockchain technology can significantly enhance the standardization of the higher education management system by providing a secure, decentralized, and transparent platform for managing educational records and seamless communication between different educational institutions in a common standard. Interoperability and compatibility of different systems, as well as accessibility and regulation, are critical factors to be addressed to make the technology more widespread among educational institutions \cite{b26}. For example, in the healthcare sector, the Health Insurance Portability and Accountability Act (HIPAA) sets standards for the privacy and security of health information \cite{HIPPA}, \cite{saha2023analysis}. Similarly, in the higher education management sector, the use of blockchain's decentralized features can help in the standardization of processes in Learning Content Management Systems (LCMS) \cite{LCMS}, Learning Record Stores (LRS) \cite{LR}, and Learning Resource Metadata \cite{meta}.

The integration of blockchain technology with the higher education management system can provide a transparent, secure, and compliant platform, overcoming the regulatory challenges faced by the current system \cite{b37}. It can provide a secure and transparent platform for data management, overcoming the issues of data disregard and data mismanagement. With the use of smart contracts, blockchain can also automate and enforce regulations, providing adequate smart contract management and ensuring compliance. Additionally, blockchain can provide a decentralized and immutable record of all transactions, making auditing \cite{auditing} more efficient and transparent, and overcoming the issue of inadequate auditing. Moreover, blockchain technology can enhance the cybersecurity  \cite{b50} of the higher education management system by providing an immutable and secure platform for data storage and communication. By eliminating the need for intermediaries, blockchain can reduce legal vagueness and provide a standardized platform for integration.

 Mohammad et al. \cite{usecase1} shed light on the significance of various use cases for the real-world implementation of blockchain technology in higher education management systems, such as learning analytics, adaptive learning, personalized learning, and collaborative learning. These use cases highlight the potential of blockchain to provide a secure, decentralized, and transparent platform for facilitating collaborations, automating administrative processes and improving interoperability, thereby enhancing the trust and adoption of the system. Integrating blockchain technology in higher education management can potentially reduce costs and provide a clear ROI \cite{roi}. Decentralization reduces infrastructure costs \cite{b13}, while automation reduces development, training, and maintenance costs. The transparent nature of blockchain reduces auditing and compliance costs. Streamlined processes and reduced costs provide a clear ROI over time.

Blockchain integration can improve scalability in the higher education management system by providing a decentralized network that can handle large volumes of data, reducing latency and improving throughput. It increases data storage capacity and reduces complexity, making it easier to manage and process data. Blockchain's consensus mechanism eliminates intermediaries and reduces transactional overhead, potentially enhancing system efficiency. Blockchain technology can bring many benefits to the higher education management system, including improved security, transparency, and efficiency. It can also address concerns related to privacy, standardization, regulatory compliance, and cost while enhancing scalability.

\subsection{Assessing Potential Benefits and Measuring Impact on Efficiency, Security, and Transparency (RQ2)}
\begin{figure}[t!]
\centerline{\includegraphics[scale=0.37]{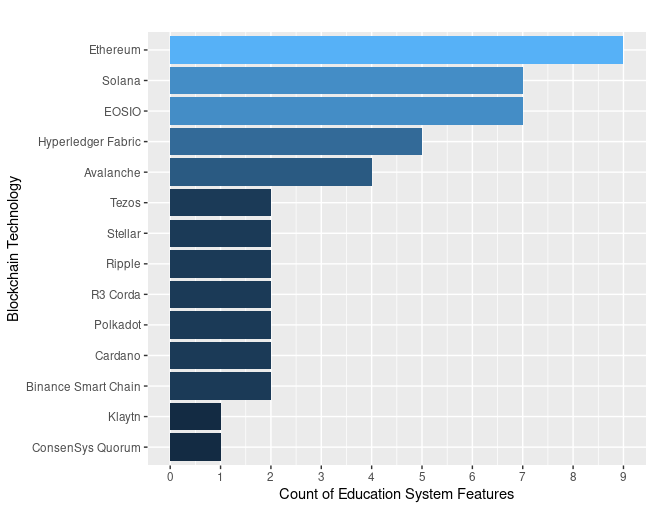}}
\caption{Blockchain Technology Solutions for Features in higher education management system}
\label{fig:RQ2}
\vspace{-4mm}
\end{figure}

Blockchain technology exhibits its capacity to improve various aspects of the higher education management system, including records management, authentication, security, privacy, and access management. Table \ref{tab:RQ2} shows the feature-wise advantages of the implementation of blockchain in the higher education management system and available supported blockchain platforms. One of the key benefits is the provision of immutable records. The utilization of different types of blockchain-based technology  for instance Ethereum, ConsenSys Quorum \cite{ConsenSys}, EOSIO \cite{eosio}, Avalanche \cite{avalanche}, Cardano \cite{Cardano}, Hyperledger Fabric \cite{hyperledger_fabric}, R3 Corda \cite{R3_Corda}, Solana \cite{solana}, Tezos \cite{tezos}, Polkadot \cite{polka} guarantees that records cannot be altered, thus providing a secure and dependable method of storing and validating student information \cite{b68}. The coverage of different higher education system features by various blockchain technologies can be visualized in Figure \ref{fig:RQ2}. The use of blockchain technology can improve accreditation and certification processes by guaranteeing the genuineness and accuracy of issued credentials. The elimination of a central authority due to the decentralized nature of blockchain technology increases transparency and trustworthiness. By enabling interoperable solutions such as Ethereum-based student record management \cite{b12}, blockchain technology can enhance record sharing across various platforms, improving the efficiency and effectiveness of record management. This is of utmost importance for recognizing prior learning and lifelong learning.

\begin{figure}[b!]
\centerline{\includegraphics[scale=0.40]{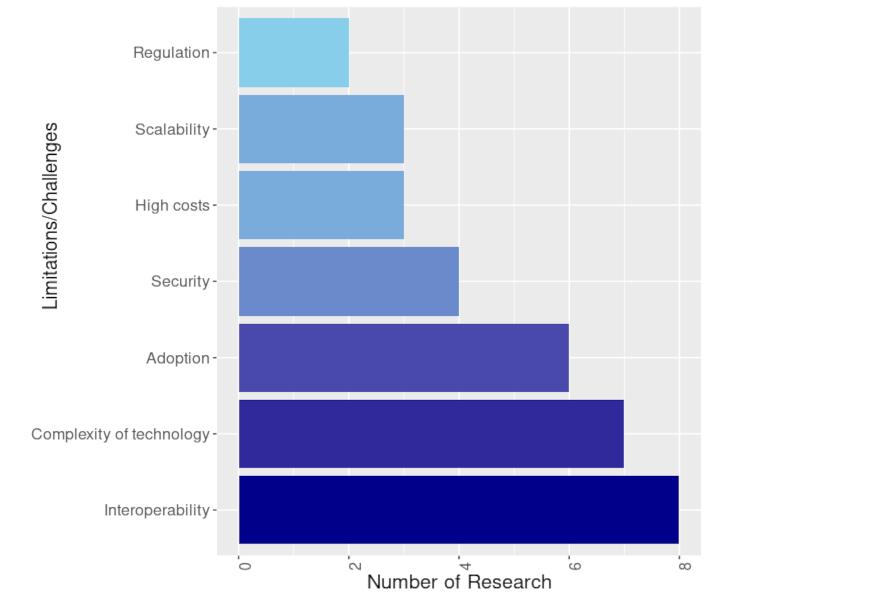}}
\caption{Research Focus on Limitations/Challenges in Implementing Blockchain in Higher Education Management System}
\label{fig:RQ3}
\end{figure}

Blockchain technology can provide cost-effective solutions in the management of higher education, by reducing administrative costs. The utilization of Hyperledger blockchain for learning management systems has been demonstrated to ensure secure storage of educational content while being cost-effective, as evidenced by studies conducted by the authors of papers \cite{b71} and \cite{b72}. This can be particularly beneficial for educational institutions operating with limited resources and facing budget constraints.

\begin{table*}[ht!]
\footnotesize
\centering
\setlength{\tabcolsep}{10pt}
  \begin{tabularx}{\textwidth}{>{\hsize=.2\hsize\centering}X|>{\hsize=.30\hsize}X|>{\hsize=.30\hsize}X|>{\hsize=.2\hsize}X}
\hline
\textbf{Higher Education Management System Features} &
  \textbf{Measured by Technology} &
  \textbf{Beneficial Outcomes} &
  \textbf{References} \\ \hline
Digital Credentials &
  \begin{minipage}{4.6cm}
   Ethereum, ConsenSys Quorum, EOSIO, Avalanche, Cardano, Hyperledger Fabric, R3 Corda, Solana, Tezos   
  \end{minipage}
 &
\begin{minipage}{4.6cm}
\begin{itemize}
 \vspace{3pt} 
 \item  Immutable Records
     \item Decentralized Verification
      \item Secure
      \item Portable
 \vspace{3pt} 
\end{itemize}
\end{minipage}
&
\cite{b71},\cite{b54},\cite{avalanche},\cite{Cardano},
\cite{eosio},\cite{hyperledger_fabric},\cite{R3_Corda},\cite{solana},\cite{tezos}
 \\ \hline
Student Records Management &
  \begin{minipage}{4.6cm}
   Ethereum, Avalanche, EOSIO, Hyperledger Fabric, R3 Corda, Solana   
  \end{minipage} &
\begin{minipage}{4.6cm}
\begin{itemize}
 \vspace{3pt} 
 \item  Interoperability
     \item Cost-effective
 \vspace{3pt} 
\end{itemize}
\end{minipage} 
  &
  \cite{b12},\cite{avalanche},\cite{eosio},\cite{hyperledger_fabric},
  \cite{R3_Corda},\cite{solana}
\\ \hline
Micro-credentials &
  \begin{minipage}{4.6cm}
   Ethereum,  Avalanche,  Cardano,  ConsenSys Quorum,  EOSIO,  Hyperledger Fabric,  R3 Corda,  Solana,  Tezos   
  \end{minipage} &

\begin{minipage}{4.6cm}
\begin{itemize}
 \vspace{3pt} 
 \item  Secure and Transparent
     \item Issuance of Credentials
      \item Secure Transparent Verification of Credentials
 \vspace{3pt} 
\end{itemize}
\end{minipage} 
 &
\cite{avalanche},\cite{Cardano},\cite{eosio},\cite{hyperledger_fabric},
\cite{R3_Corda},\cite{solana},\cite{tezos},\cite{b71} \\
 \hline
Learning Management Systems &
  \begin{minipage}{4.6cm}
   Avalanche, Hyperledger Fabric, EOSIO, Polkadot, R3 Corda, Solana, Tezos  
  \end{minipage} &
  \begin{minipage}{4.6cm}
\begin{itemize}
 \vspace{3pt} 
 \item  Decentralized Tracking \& Recording of Progress
     \item Secure Storage of Educational Content
      \item Cost-effective
 \vspace{3pt} 
\end{itemize}
\end{minipage}
   &
   \cite{b54},\cite{avalanche},\cite{Cardano},\cite{eosio},
   \cite{hyperledger_fabric},\cite{R3_Corda},\cite{solana},\cite{tezos}\\
  \hline
Payment Systems &
  \begin{minipage}{4.6cm}
   \vspace{3pt} 
   Etherium, Binance Smart Chain, EOSIO, Hyperledger Fabric, Polkadot, R3 Corda, Ripple, Stellar   
   \vspace{3pt} 
  \end{minipage} &
\begin{minipage}{4.6cm}
\begin{itemize}
 \vspace{3pt} 
 \item  Secure Transaction
     \item Transparent Transaction
 \vspace{3pt} 
\end{itemize}
\end{minipage} &
\cite{b20},\cite{b54},\cite{avalanche},\cite{Cardano},
\cite{eosio},\cite{hyperledger_fabric},\cite{R3_Corda},\cite{solana},
\cite{b60}\\
   \hline
Digital Identity &
  \begin{minipage}{4.6cm}
   \vspace{3pt} 
   Ethereum, Avalanche, Cardano, ConsenSys Quorum, EOSIO, Hyperledger Fabric, R3 Corda, Solana, Tezos  
    \vspace{3pt} 
  \end{minipage} &
    \begin{minipage}{4.6cm}
\begin{itemize}
 \vspace{3pt} 
 \item  Identity Verification
     \item Secure Storage of Personal Information
 \vspace{3pt} 
\end{itemize}
\end{minipage}
 &
  \cite{b46},\cite{avalanche},\cite{Cardano},\cite{eosio},
  \cite{hyperledger_fabric},\cite{R3_Corda},\cite{solana},\cite{tezos}\\
 \hline
Student Loan Management &
  \begin{minipage}{4.6cm}
   Etherium, Binance Smart Chain, EOSIO, Hyperledger Fabric, Polkadot, R3 Corda, Ripple, Stellar   
  \end{minipage} &
      \begin{minipage}{4.6cm}
\begin{itemize}
 \vspace{3pt} 
 \item  Secure and Transparent Loan Issuance
     \item Secure and Transparent Loan Management
      \item Secure and Transparent Loan Repayment
 \vspace{3pt} 
\end{itemize}
\end{minipage}
  &
  \cite{b73},\cite{b54},\cite{b55},\cite{b56},
  \cite{b57},\cite{b58},\cite{b59},\cite{b60},\cite{b61}
  \\ \hline
Online Learning &
  \begin{minipage}{4.6cm}
   Ethereum, Avalanche, EOSIO, Hyperledger Sawtooth, Klaytn,  Polkadot, R3 Corda, Solana, Tezos   
  \end{minipage} &
        \begin{minipage}{4.6cm}
\begin{itemize}
 \vspace{3pt} 
 \item  Decentralized Content Sharing
     \item Secure Tracking of Student Progress
      \item Portable Student Records
 \vspace{3pt} 
\end{itemize}
\end{minipage}
 &
 \cite{b46},\cite{avalanche},\cite{Cardano},\cite{eosio},
 \cite{hyperledger_fabric},\cite{R3_Corda},\cite{solana},\cite{tezos}\\
  \hline
Data Analytics &
  \begin{minipage}{4.6cm}
   Blockcerts, Avalanche, EOSIO, Hyperledger Sawtooth, R3 Corda, Solana  
  \end{minipage} &
          \begin{minipage}{4.6cm}
\begin{itemize}
 \vspace{3pt} 
 \item  Secure and Transparent Data Collection
     \item Secure and Transparent Data Storage
      \item Secure and Transparent Data Analysis
 \vspace{3pt} 
\end{itemize}
\end{minipage}
   &
   \cite{b37},\cite{80},\cite{avalanche},\cite{Cardano},
   \cite{eosio},\cite{R3_Corda},\cite{solana},\cite{b61}\\ 
   \hline
Peer-reviewing Process &
  \begin{minipage}{4.6cm}
   Ethereum, Avalanche, Cardano, EOSIO, Hyperledger Fabric, R3 Corda, Solana   
  \end{minipage} &
            \begin{minipage}{4.6cm}
\begin{itemize}
 \vspace{3pt} 
 \item  Secure and Transparent Submission of Papers
     \item Selection of Reviewers
      \item Review Results
 \vspace{3pt} 
\end{itemize}
\end{minipage}
 &
\cite{avalanche},\cite{Cardano},\cite{eosio},\cite{hyperledger_fabric},
\cite{R3_Corda},\cite{solana}
 \\ \hline
Anti-plagiarism &
  \begin{minipage}{4.6cm}
   Ethereum, Avalanche, Cardano, EOSIO, Hyperledger Fabric, R3 Corda, Solana  
  \end{minipage} &
              \begin{minipage}{4.6cm}
\begin{itemize}
 \vspace{3pt} 
 \item  Tamper-proof Tracking of Written Work
     \item Verification of Originality
      \item Secure Storage of Written Work
 \vspace{3pt} 
\end{itemize}
\end{minipage}
 &
 \cite{b75},\cite{avalanche},\cite{Cardano},\cite{eosio},
 \cite{hyperledger_fabric},\cite{R3_Corda},\cite{solana},\cite{b60},\cite{b61}
  \\ \hline
\end{tabularx}%
\caption{Present blockchain applications in higher education management ecosystem}
\label{tab:RQ2}
\vspace{-4mm}
\end{table*}

Skiba et al. \cite{b20} have demonstrated the potential of blockchain technology to offer secure and transparent transactions in higher education management systems through the utilization of Ripple blockchain for payment systems. This can enhance the trustworthiness and efficiency of financial transactions in the sector, especially for tuition fees, scholarships, and student loan management \cite{b73}. Further benefits of blockchain technology in this domain includes security and transparency, interoperability, and cost-effectiveness of the solutions, as well as the overall improvement in the learning experience. Another significant aspect of higher education management that Palmisano et al. \cite{plagiarism} explored is the utilization of blockchain technology for implementing anti-plagiarism measures. This can promote original thought and creativity while ensuring tamper-proof tracking and verification of written work \cite{b75}.

\begin{table*}[ht!]
\footnotesize
\centering
\setlength{\tabcolsep}{0.1em}
 \begin{tabularx}{\textwidth}{>{\hsize=.18\hsize\centering\arraybackslash}X|>{\hsize=.18\hsize\centering\arraybackslash}X|>{\hsize=.28\hsize\centering\arraybackslash}X|>{\hsize=.28\hsize\centering\arraybackslash}X|>{\hsize=.08\hsize\centering\arraybackslash}X}
\hline
\textbf{Limitations of Incorporating Blockchain} &
  \textbf{Challenges of Incorporating Blockchain} &
  \textbf{Research Idea} &
  \textbf{Potential Solutions} &
  \textbf{References} \\ \hline
Complexity of Technology &
  Difficulty in understanding and implementing &
  \begin{minipage}{4.8cm}
  \vspace{3pt}
  \begin{itemize}
      \item Conduct feasibility studies and cost-benefit analysis of blockchain implementation in educational institutions
      \item Investigate the technical expertise and training needs of educators and institutions
  \end{itemize}
  \vspace{3pt}
  \end{minipage} &
  \begin{minipage}{4.8cm}
  \vspace{3pt}
   \begin{itemize}
      \item Offer pedagogical and professional development programs for institutions and educators  
      \item Develop easy-to-use blockchain platforms and tools for educational institutions
      \item Minimize implementation cost by utilizing existing infrastructure and resources
    \end{itemize}
    \vspace{3pt}
   \end{minipage} &
   \cite{b37},\cite{b19},
   \cite{b73},\cite{b71},
   \cite{b75},\cite{76}\\
    \hline
Interoperability &
  Difficulty in integrating with existing systems &
  \begin{minipage}{4.8cm}
  \begin{itemize}
      \item Conduct technical compatibility assessments
      \item Investigate standardization efforts and regulatory compliance requirements
      \item Analyze current interoperability solutions for educational systems
    \end{itemize}
  \end{minipage} &
  \begin{minipage}{4.8cm}
  \vspace{3pt}
  \begin{itemize}
      \item Develop standardized protocols and interfaces
      \item Work with government agencies and stakeholders to establish regulations 
      \item Introduce interoperability solutions that work with current systems
      \end{itemize}
      \vspace{3pt}
      \end{minipage} &
      \cite{77},\cite{78},
      \cite{79},\cite{b34},
      \cite{b63},\cite{80},
      \cite{81},\cite{82}\\
      \hline
Scalability &
  Difficulty in handling large number of users and transactions &
  \begin{minipage}{4.8cm}
  \vspace{3pt}
  \begin{itemize}
      \item Investigate scalability solutions such as sharding and off-chain scaling
      \item Study the infrastructure requirements for large-scale blockchain implementation in the higher education management sector
        \end{itemize}
        \vspace{3pt}
      \end{minipage} &
  \begin{minipage}{4.8cm}
  \begin{itemize}
      \item Implement scalability solutions such as sharding and off-chain scaling
       \item Leverage existing infrastructure and resources to reduce costs
       \end{itemize}
       \end{minipage} &
        \cite{b54},\cite{b60},
        \cite{80},
        \\ \hline
Security &
  Risk of hacking and data breaches &
  \begin{minipage}{4.8cm}
  \begin{itemize}
      \item Conduct threat modeling and risk assessments 
      \item Investigate security solutions for educational institutions and their data
        \end{itemize}
      \end{minipage} &
  \begin{minipage}{4.8cm}
  \vspace{3pt}
  \begin{itemize}
      \item Implement robust security measures such as encryption and multi-factor authentication 
      \item Provide training and education programs on security best practices
        \end{itemize}
        \vspace{3pt}
      \end{minipage} &
      \cite{b12},\cite{b15},
      \cite{b27},\cite{b28},
      \cite{81},\cite{83},
      \cite{84}
   \\ \hline
Regulation &
  Uncertainty surrounding government regulations and legal issues &
  \begin{minipage}{4.8cm}
  \vspace{3pt}
  \begin{itemize}
      \item Study the regulatory and legal frameworks surrounding blockchain technology in the higher education management sector \item Investigate the compliance requirements for educational institutions
       \end{itemize}
       \vspace{3pt}
      \end{minipage} &
  \begin{minipage}{4.8cm}
  \vspace{3pt}
  \begin{itemize}
      \item Collaborate with government agencies and other stakeholders to establish clear regulations and guidelines
      \item Develop standard operating procedures for compliance
      \end{itemize}
      \vspace{3pt}
      \end{minipage} &
      \cite{b50},\cite{b58}, \cite{randolph2022blockchain}
   \\ \hline
High Costs &
  High costs associated with the implementation and maintenance &
  \begin{minipage}{4.8cm}
  \begin{itemize}
      \item Conduct cost-benefit analysis of blockchain implementation in the higher education management sector 
      \item Investigate cost-effective solutions such as open-source blockchain platforms and cloud-based infrastructure
      \end{itemize}
      \end{minipage} &
  \begin{minipage}{4.8cm}
  \vspace{3pt}
  \begin{itemize}
      \item Adopt cost-efficient solutions (open-source blockchain platforms and cloud-based infrastructure) to lower expenses 
      \item Create a funding and revenue model to support the implementation and upkeep of blockchain in higher education management system
      \end{itemize}
      \vspace{3pt}
      \end{minipage} &
      \cite{b13},\cite{b14},
      \cite{b39}
   \\ \hline
Adoption &
  Limited number of institutions using blockchain technology &
  \begin{minipage}{4.8cm}
  \vspace{3pt}
  \begin{itemize}
      \item Investigate barriers to the adoption of blockchain in higher education management system 
      \item Assess the awareness and understanding of the benefits of blockchain in education
      \end{itemize}
      \vspace{3pt}
      \end{minipage} &
      \begin{minipage}{4.8cm}
      \vspace{3pt}
      \begin{itemize}
      \item Increase awareness and understanding of the potential benefits of blockchain technology in the higher education management sector, develop
      \end{itemize}
      \vspace{3pt}
      \end{minipage}
      &
     \cite{b9},\cite{b10},
     \cite{b27},\cite{b54},
     \cite{b55},\cite{b63},
     \cite{b71}, \cite{faruk2022development}
   \\ \hline
\end{tabularx}%
\caption{Limitations,challenges and solutoins of implementing blockchain technology in the higher education management sector}
\label{tab:RQ3}
\vspace{-4mm}
\end{table*}

\vspace{-1mm}

Furthermore, Blockchain enables issuance and verification of micro-credentials, useful for skills-based learning and informal recognition \cite{b71}. Alam et al. \cite{b42} accentuated that by utilizing Ethereum, the credentials can be secured and made transparent, thereby improving the credibility of the system. It also enables secure, decentralized tracking and recording of student progress in online education systems  \cite{b33}.
Overall, blockchain technology offers a range of advantages for the higher education management sector, including secure record-keeping, improved authentication and access management, and cost-effective solutions, as evidenced by various studies. Additionally, the technology provides transparent and trustworthy transactions, enhanced accreditation processes, and anti-plagiarism measures, thereby improving the overall learning experience.
\subsection{Challenges and Solutions of Blockchain in Higher Education Management system (RQ3)}
The implementation of blockchain technology in the educational sector is still in its early stages, but it has the potential to fundamentally revolutionize how higher education is managed and delivered \cite{b56}. Despite its advantages, implementing blockchain technology in the higher education management sector faces a number of challenges that must be surmounted in order to realize all of its potential \cite{b53}. Figure \ref{fig:RQ3} illustrates the research's emphasis on the difficulties and restrictions of integrating blockchain technology into higher education management systems.
The inherent technological complexity of blockchain technology, according to Yumna et al. \cite{b3}, is one of the main barriers to its effective integration in the higher education management sector. For institutions and educators, it may be challenging to comprehend and use blockchain technology effectively due to its complexity. For the successful implementation of blockchain systems in education, Anwar et al. \cite{b58} highlighted the potential challenge of technical knowledge gaps and challenges in navigating the rapidly changing blockchain technology landscape.

A cost-benefit analysis and feasibility study of integrating blockchain in educational institutions are required to get around this problem. The most effective and efficient ways to implement blockchain technology in the higher education management sector should be determined by these studies, which should also look at the technical expertise and training requirements of educators and institutions \cite{b31}. Additionally, by giving educators and institutions the abilities and knowledge necessary to successfully integrate blockchain technology into their operations, pedagogical training and professional development programs for institutions and educators can help to alleviate this challenge \cite{peda}.

Fedorova et al. \cite{b39}, \cite{b16}, Raimundo et al. \cite{b12}, and Ghaffar et al. \cite{b35}, have identified a significant barrier to the adoption of blockchain technology in higher education management systems as limited interoperability. This difficulty makes it difficult for blockchain systems to be integrated with current infrastructure, which results in a lack of standardization and compliance. The authors suggest a number of approaches to resolve this problem, including technical compatibility evaluations, standardization initiatives, regulatory compliance analysis, application of standardized protocols, cooperation with governmental organizations, and integration of interoperability solutions with current systems. By creating open source technologies that work with a variety of platforms and systems, the Hyperledger project of the Linux Foundation seeks to address the issue of the blockchain sector's poor interoperability. This collaboration between multiple organizations is aimed at establishing standardized protocols and interfaces, leading to greater compatibility in the industry \cite{78}, \cite{79}, \cite{82}.

Scalability is another issue with blockchain technology in the higher education management field. Although blockchain technology has the capacity to support numerous users and transactions, scaling it to meet the needs of a user base that is expanding quickly can be challenging \cite{b54}. Due to this, educational institutions may find it challenging to take full advantage of the potential advantages of blockchain technology, such as improved efficiency and lower costs.

Institutions must carefully consider the scalability of the blockchain technology they choose to implement in order to overcome this difficulty. This entails carrying out technical analyses of the technology's scalability and assessing its capacity to handle the anticipated volume of transactions and data \cite{b60}. Additionally, organizations ought to think about implementing alternative blockchain solutions that can help the technology scale, like sharding or off-chain transactions \cite{80}.

The final challenge for blockchain technology in higher education management systems is security. There are still dangers associated with using blockchain technology, such as hacking, data breaches, and other security threats \cite{81}, despite its many benefits, such as secure data management and decentralized systems. Sensitive data may be lost as a result, and educational institutions' reputations may suffer. The adoption of blockchain technology in student financial aid poses a security challenge because it increases the likelihood of data breaches. To mitigate this, robust security measures such as encryption and multi-factor authentication must be implemented to protect sensitive information and ensure secure operations.\cite{83}, \cite{84}. Table \ref{tab:RQ3} provided a comprehensive overview of the multifaceted challenges that arise in the implementation of blockchain technology in higher education management systems, and also offered a range of potential solutions to overcome these limitations.
 
\subsection{Our Findings and Shaping Relevant Debates}
The results of the analysis suggest that blockchain technology has significant potential to address privacy and security concerns in higher education management systems. According to the visualization in Figure \ref{fig:RQ1}, privacy and security emerges as the foremost domain for implementing blockchain technology, underscoring its potential to address critical challenges in safeguarding sensitive data and preventing unauthorized access. Based on Figure \ref{fig:RQ2}, the use of Ethereum is recommended for implementing a comprehensive higher education management system, given its potential applicability across all areas of such a system. Figure \ref{fig:RQ3} points out the scarcity of research on regulatory and scalability considerations in implementing blockchain in higher education management systems, emphasizing the need for future research to prioritize these critical areas. As the technology continues to evolve and mature, it is important to continue exploring its potential to improve standardization and enhance the overall quality of higher education management.

\section{Conclusion}
Blockchain technology offers a secure and immutable way to store and access educational credentials and records, with the potential to revolutionize the higher education management sector. By reducing the risk of fraud and providing a tamper-proof record, blockchain technology can improve the efficiency of the higher education management system. Despite challenges such as lack of expertise and slow processes, the research shows the benefits of blockchain in higher education management system and the need to address these challenges.

\section*{Acknowledgment}
\footnotesize
This work is partially supported by National Science Foundation (NSF) under awards \#2100115.

\renewcommand{\bibfont}{\footnotesize}

\bibliographystyle{IEEEtranN}
\bibliography{biblography.bib}

% Generated by IEEEtranN.bst, version: 1.14 (2015/08/26)
\begin{thebibliography}{96}
\providecommand{\natexlab}[1]{#1}
\providecommand{\url}[1]{#1}
\csname url@samestyle\endcsname
\providecommand{\newblock}{\relax}
\providecommand{\bibinfo}[2]{#2}
\providecommand{\BIBentrySTDinterwordspacing}{\spaceskip=0pt\relax}
\providecommand{\BIBentryALTinterwordstretchfactor}{4}
\providecommand{\BIBentryALTinterwordspacing}{\spaceskip=\fontdimen2\font plus
\BIBentryALTinterwordstretchfactor\fontdimen3\font minus
  \fontdimen4\font\relax}
\providecommand{\BIBforeignlanguage}[2]{{%
\expandafter\ifx\csname l@#1\endcsname\relax
\typeout{** WARNING: IEEEtranN.bst: No hyphenation pattern has been}%
\typeout{** loaded for the language `#1'. Using the pattern for}%
\typeout{** the default language instead.}%
\else
\language=\csname l@#1\endcsname
\fi
#2}}
\providecommand{\BIBdecl}{\relax}
\BIBdecl

\bibitem[Wei(2022)]{stats}
D.~Wei, ``Gemiverse: The blockchain-based professional certification and
  tourism platform with its own ecosystem in the metaverse,''
  \emph{International Journal of Geoheritage and Parks}, vol.~10, no.~2, pp.
  322--336, 2022.

\bibitem[Raman(2019)]{stats2}
A.~Raman, ``Potentials of fog computing in higher education.''
  \emph{International Journal of Emerging Technologies in Learning}, vol.~14,
  no.~18, 2019.

\bibitem[Hackman and Reindl(2022)]{stats3}
S.~T. Hackman and S.~Reindl, ``Challenging edtech: Towards a more inclusive,
  accessible and purposeful version of edtech,'' \emph{Knowledge Cultures},
  vol.~10, no.~1, pp. 7--21, 2022.

\bibitem[Chen et~al.(2022)Chen, Bohloul, Ma, and Li]{stats4}
W.~Chen, S.~M. Bohloul, Y.~Ma, and L.~Li, ``A blockchain-based information
  management system for academic institutions: a case study of international
  students’ workflow,'' \emph{Information Discovery and Delivery}, vol.~50,
  no.~4, pp. 343--352, 2022.

\bibitem[Panagiotidis(2022)]{stats5}
P.~Panagiotidis, ``Blockchain in education-the case of language learning,''
  \emph{European Journal of Education}, vol.~5, no.~1, pp. 66--82, 2022.

\bibitem[Chowdhury et~al.(2022)Chowdhury, Rodriguez-Espindola, Dey, and
  Budhwar]{state6}
S.~Chowdhury, O.~Rodriguez-Espindola, P.~Dey, and P.~Budhwar, ``Blockchain
  technology adoption for managing risks in operations and supply chain
  management: evidence from the uk,'' \emph{Annals of operations research}, pp.
  1--36, 2022.

\bibitem[Yu and Sannikova(2021)]{state7}
K.~Yu and L.~Sannikova, ``Digital platforms in china and europe: Legal
  challenges,'' \emph{BRICS Law Journal}, vol.~8, no.~3, pp. 121--147, 2021.

\bibitem[Mardisentosa et~al.(2021)Mardisentosa, Rahardja, Zelina, Oganda, and
  Hardini]{state8}
B.~Mardisentosa, U.~Rahardja, K.~Zelina, F.~P. Oganda, and M.~Hardini,
  ``Sustainable learning micro-credential using blockchain for student
  achievement records,'' in \emph{2021 Sixth International Conference on
  Informatics and Computing (ICIC)}.\hskip 1em plus 0.5em minus 0.4em\relax
  IEEE, 2021, pp. 1--6.

\bibitem[Al~Hilali and Shaker(2021)]{state9}
R.~A. Al~Hilali and H.~Shaker, ``Blockchain technology's status of
  implementation in oman: Empirical study,'' \emph{International Journal of
  Computing and Digital Systems}, 2021.

\bibitem[Wolz et~al.(2021)Wolz, Gottlieb, and Pongratz]{state10}
E.~Wolz, M.~Gottlieb, and H.~Pongratz, ``Digital credentials in higher
  education institutions: A literature review,'' \emph{Innovation Through
  Information Systems: Volume III: A Collection of Latest Research on
  Management Issues}, pp. 125--140, 2021.

\bibitem[Al~Harthy et~al.(2019{\natexlab{a}})Al~Harthy, Al~Shuhaimi, and
  Al~Ismaily]{b10}
K.~Al~Harthy, F.~Al~Shuhaimi, and K.~K.~J. Al~Ismaily, ``The upcoming
  blockchain adoption in higher-education: requirements and process,'' in
  \emph{2019 4th MEC international conference on big data and smart city
  (ICBDSC)}.\hskip 1em plus 0.5em minus 0.4em\relax IEEE, 2019, pp. 1--5.

\bibitem[Ma and Fang(2020)]{b40}
Y.~Ma and Y.~Fang, ``Current status, issues, and challenges of blockchain
  applications in education,'' \emph{International Journal of Emerging
  Technologies in Learning (IJET)}, vol.~15, no.~12, pp. 20--31, 2020.

\bibitem[Sharples and Domingue(2016)]{b19}
M.~Sharples and J.~Domingue, ``The blockchain and kudos: A distributed system
  for educational record, reputation and reward,'' in \emph{European conference
  on technology enhanced learning}.\hskip 1em plus 0.5em minus 0.4em\relax
  Springer, 2016, pp. 490--496.

\bibitem[Skiba et~al.(2017)]{b20}
D.~J. Skiba \emph{et~al.}, ``The potential of blockchain in education and
  health care,'' \emph{Nursing education perspectives}, vol.~38, no.~4, pp.
  220--221, 2017.

\bibitem[Rooksby and Dimitrov(2017)]{b21}
J.~Rooksby and K.~Dimitrov, ``Trustless education? a blockchain system for
  university grades,'' in \emph{New Value Transactions: Understanding and
  Designing for Distributed Autonomous Organisations, Workshop at DIS}, 2017.

\bibitem[Hardini et~al.(2020)Hardini, Aini, Rahardja, Izzaty, and
  Faturahman]{b13}
M.~Hardini, Q.~Aini, U.~Rahardja, R.~D. Izzaty, and A.~Faturahman, ``Ontology
  of education using blockchain: Time based protocol,'' in \emph{2020 2nd
  International Conference on Cybernetics and Intelligent System
  (ICORIS)}.\hskip 1em plus 0.5em minus 0.4em\relax IEEE, 2020, pp. 1--5.

\bibitem[Kuvshinov et~al.(2018{\natexlab{a}})Kuvshinov, Nikiforov, Mostovoy,
  Mukhutdinov, Andreev, and Podtelkin]{b24}
K.~Kuvshinov, I.~Nikiforov, J.~Mostovoy, D.~Mukhutdinov, K.~Andreev, and
  V.~Podtelkin, ``Disciplina: Blockchain for education,'' \emph{Yellow Paper.
  URL: https://disciplina. io/yellowpaper. pdf}, 2018.

\bibitem[Rahardja et~al.(2020)Rahardja, Aini, Ngadi, Hardini, and Oganda]{b9}
U.~Rahardja, Q.~Aini, M.~A. Ngadi, M.~Hardini, and F.~P. Oganda, ``The
  blockchain manifesto,'' in \emph{2020 2nd International Conference on
  Cybernetics and Intelligent System (ICORIS)}.\hskip 1em plus 0.5em minus
  0.4em\relax IEEE, 2020, pp. 1--5.

\bibitem[Anwar et~al.(2022)Anwar, Rahardja, Prawiyogi, Santoso, and
  Maulana]{b58}
A.~S. Anwar, U.~Rahardja, A.~G. Prawiyogi, N.~P.~L. Santoso, and S.~Maulana,
  ``ilearning model approach in creating blockchain based higher education
  trust,'' \emph{Int. J. Artif. Intell. Res}, vol.~6, no.~1, 2022.

\bibitem[Aini et~al.(2021{\natexlab{a}})Aini, Lutfiani, Santoso, Sulistiawati,
  and Astriyani]{b47}
Q.~Aini, N.~Lutfiani, N.~P.~L. Santoso, S.~Sulistiawati, and E.~Astriyani,
  ``Blockchain for education purpose: essential topology,'' \emph{Aptisi
  Transactions on Management (ATM)}, vol.~5, no.~2, pp. 112--120, 2021.

\bibitem[Aini et~al.(2021{\natexlab{b}})Aini, Lutfiani, Santoso, Sulistiawati,
  and Astriyani]{b6}
------, ``Blockchain for education purpose: essential topology,'' \emph{Aptisi
  Transactions on Management (ATM)}, vol.~5, no.~2, pp. 112--120, 2021.

\bibitem[Chen et~al.(2018)Chen, Xu, Lu, and Chen]{b8}
G.~Chen, B.~Xu, M.~Lu, and N.-S. Chen, ``Exploring blockchain technology and
  its potential applications for education,'' \emph{Smart Learning
  Environments}, vol.~5, no.~1, pp. 1--10, 2018.

\bibitem[Han et~al.(2018{\natexlab{a}})Han, Li, He, Wu, Xie, and Baba]{b4}
M.~Han, Z.~Li, J.~He, D.~Wu, Y.~Xie, and A.~Baba, ``A novel blockchain-based
  education records verification solution,'' in \emph{Proceedings of the 19th
  annual SIG conference on information technology education}, 2018, pp.
  178--183.

\bibitem[Lutfiani et~al.(2021)Lutfiani, Aini, Rahardja, Wijayanti, Nabila, and
  Ali]{b11}
N.~Lutfiani, Q.~Aini, U.~Rahardja, L.~Wijayanti, E.~A. Nabila, and M.~I. Ali,
  ``Transformation of blockchain and opportunities for education 4.0,''
  \emph{International Journal of Education and Learning}, vol.~3, no.~3, pp.
  222--231, 2021.

\bibitem[Wang et~al.(2019)Wang, Zhang, Xiao, Chung, and Cai]{b14}
G.~Wang, H.~Zhang, B.~Xiao, Y.-C. Chung, and W.~Cai, ``Edubloud: A
  blockchain-based education cloud,'' in \emph{2019 Computing, Communications
  and IoT Applications (ComComAp)}.\hskip 1em plus 0.5em minus 0.4em\relax
  IEEE, 2019, pp. 352--357.

\bibitem[Amitkumar and Apriliasari(2021)]{b55}
M.~Amitkumar and D.~Apriliasari, ``Blockchain technology application:
  Authentication system in digital education,'' 2021.

\bibitem[Kumar et~al.(2021)Kumar, Verma, Gupta, and Kumar]{b57}
R.~Kumar, A.~Verma, B.~Gupta, and S.~Kumar, ``Dual-tree sparse decomposition of
  dwt filters for ecg signal compression and hrv analysis,'' \emph{Augmented
  Human Research}, vol.~6, no.~1, pp. 1--7, 2021.

\bibitem[Belen-Saglam et~al.(2023)Belen-Saglam, Altuncu, Lu, and Li]{b61}
R.~Belen-Saglam, E.~Altuncu, Y.~Lu, and S.~Li, ``A systematic literature review
  of the tension between the gdpr and public blockchain systems,''
  \emph{Blockchain: Research and Applications}, p. 100129, 2023.

\bibitem[Yang et~al.(2023)Yang, Shen, Zhong, and Liao]{b59}
L.~Yang, X.~Shen, C.~Zhong, and Y.~Liao, ``On-demand inference acceleration for
  directed acyclic graph neural networks over edge-cloud collaboration,''
  \emph{Journal of Parallel and Distributed Computing}, vol. 171, pp. 79--87,
  2023.

\bibitem[ROZHKOVA and OLENTSOVA(2020)]{b63}
A.~ROZHKOVA and J.~OLENTSOVA, ``Case-study method as an educational technology
  for teaching management students,'' in \emph{35th International Business
  Information Management Association Conference (IBIMA)}, 2020, pp. 3690--3697.

\bibitem[Anderson and Holloway(2020)]{b64}
K.~T. Anderson and J.~Holloway, ``Discourse analysis as theory, method, and
  epistemology in studies of education policy,'' \emph{Journal of Education
  Policy}, vol.~35, no.~2, pp. 188--221, 2020.

\bibitem[Frizzo-Barker(2021)]{discourse_analysis}
J.~Frizzo-Barker, ``Decentralizing the gender-blind meritocracy: A
  technofeminist discourse analysis of women's work in blockchain,'' 2021.

\bibitem[Hirsch and Gellner(2020)]{b65}
E.~Hirsch and D.~N. Gellner, ``Introduction: ethnography of organizations and
  organizations of ethnography,'' in \emph{Inside Organizations}.\hskip 1em
  plus 0.5em minus 0.4em\relax Routledge, 2020, pp. 1--15.

\bibitem[Ghaffar and Hussain(2019)]{b35}
A.~Ghaffar and M.~Hussain, ``Bceap-a blockchain embedded academic paradigm to
  augment legacy education through application,'' in \emph{Proceedings of the
  3rd International Conference on Future Networks and Distributed Systems},
  2019, pp. 1--11.

\bibitem[Al~Harthy et~al.(2019{\natexlab{b}})Al~Harthy, Al~Shuhaimi, and
  Al~Ismaily]{b27}
K.~Al~Harthy, F.~Al~Shuhaimi, and K.~K.~J. Al~Ismaily, ``The upcoming
  blockchain adoption in higher-education: requirements and process,'' in
  \emph{2019 4th MEC international conference on big data and smart city
  (ICBDSC)}.\hskip 1em plus 0.5em minus 0.4em\relax IEEE, 2019, pp. 1--5.

\bibitem[Kuvshinov et~al.(2018{\natexlab{b}})Kuvshinov, Nikiforov, Mostovoy,
  Mukhutdinov, Andreev, and Podtelkin]{b5}
K.~Kuvshinov, I.~Nikiforov, J.~Mostovoy, D.~Mukhutdinov, K.~Andreev, and
  V.~Podtelkin, ``Disciplina: Blockchain for education,'' \emph{Yellow Paper.
  URL: https://disciplina. io/yellowpaper. pdf}, 2018.

\bibitem[Guustaaf et~al.(2021)Guustaaf, Rahardja, Aini, Maharani, and
  Santoso]{b46}
E.~Guustaaf, U.~Rahardja, Q.~Aini, H.~W. Maharani, and N.~A. Santoso,
  ``Blockchain-based education project,'' \emph{Aptisi Transactions on
  Management (ATM)}, vol.~5, no.~1, pp. 46--61, 2021.

\bibitem[Bucea-Manea-Țoni{\c{s}} et~al.(2021)Bucea-Manea-Țoni{\c{s}},
  Martins, Bucea-Manea-Țoni{\c{s}}, Gheorghiț{\u{a}}, Kuleto, Ili{\'c}, and
  Simion]{b52}
R.~Bucea-Manea-Țoni{\c{s}}, O.~M. Martins, R.~Bucea-Manea-Țoni{\c{s}},
  C.~Gheorghiț{\u{a}}, V.~Kuleto, M.~P. Ili{\'c}, and V.-E. Simion,
  ``Blockchain technology enhances sustainable higher education,''
  \emph{Sustainability}, vol.~13, no.~22, p. 12347, 2021.

\bibitem[Delgado-von Eitzen et~al.(2021{\natexlab{a}})Delgado-von Eitzen,
  Anido-Rif{\'o}n, and Fern{\'a}ndez-Iglesias]{b50}
C.~Delgado-von Eitzen, L.~Anido-Rif{\'o}n, and M.~J. Fern{\'a}ndez-Iglesias,
  ``Application of blockchain in education: Gdpr-compliant and scalable
  certification and verification of academic information,'' \emph{Applied
  Sciences}, vol.~11, no.~10, p. 4537, 2021.

\bibitem[Kolvenbach et~al.(2018)Kolvenbach, Ruland, Gr{\"a}ther, and Prinz]{b2}
S.~Kolvenbach, R.~Ruland, W.~Gr{\"a}ther, and W.~Prinz, ``Blockchain 4
  education,'' in \emph{Proceedings of 16th European Conference on
  Computer-Supported Cooperative Work-Panels, Posters and Demos}.\hskip 1em
  plus 0.5em minus 0.4em\relax European Society for Socially Embedded
  Technologies (EUSSET), 2018.

\bibitem[Raimundo and Ros{\'a}rio(2021)]{b12}
R.~Raimundo and A.~Ros{\'a}rio, ``Blockchain system in the higher education,''
  \emph{European Journal of Investigation in Health, Psychology and Education},
  vol.~11, no.~1, pp. 276--293, 2021.

\bibitem[Fedorova and Skobleva(2020{\natexlab{a}})]{b16}
E.~P. Fedorova and E.~I. Skobleva, ``Application of blockchain technology in
  higher education.'' \emph{European Journal of Contemporary Education},
  vol.~9, no.~3, pp. 552--571, 2020.

\bibitem[Fenwick et~al.(2017)Fenwick, Kaal, and Vermeulen]{b22}
M.~Fenwick, W.~A. Kaal, and E.~P. Vermeulen, ``Legal education in the
  blockchain revolution,'' \emph{Vand. J. Ent. \& Tech. L.}, vol.~20, p. 351,
  2017.

\bibitem[Han et~al.(2018{\natexlab{b}})Han, Li, He, Wu, Xie, and Baba]{b23}
M.~Han, Z.~Li, J.~He, D.~Wu, Y.~Xie, and A.~Baba, ``A novel blockchain-based
  education records verification solution,'' in \emph{Proceedings of the 19th
  annual SIG conference on information technology education}, 2018, pp.
  178--183.

\bibitem[Sousa and de~Bem~Machado(2020)]{b41}
M.~J. Sousa and A.~de~Bem~Machado, ``Blockchain technology reshaping education:
  contributions for policy,'' in \emph{Blockchain technology applications in
  education}.\hskip 1em plus 0.5em minus 0.4em\relax IGI Global, 2020, pp.
  113--125.

\bibitem[Park(2021)]{b53}
J.~Park, ``Promises and challenges of blockchain in education,'' \emph{Smart
  Learning Environments}, vol.~8, no.~1, pp. 1--13, 2021.

\bibitem[D{\"u}dder et~al.(2021)D{\"u}dder, Fomin, G{\"u}rpinar, Henke, Iqbal,
  Janavi{\v{c}}ien{\.e}, Matulevi{\v{c}}ius, Straub, and Wu]{b56}
B.~D{\"u}dder, V.~Fomin, T.~G{\"u}rpinar, M.~Henke, M.~Iqbal,
  V.~Janavi{\v{c}}ien{\.e}, R.~Matulevi{\v{c}}ius, N.~Straub, and H.~Wu,
  ``Interdisciplinary blockchain education: Utilizing blockchain technology
  from various perspectives,'' \emph{Frontiers in Blockchain}, vol.~3, p.
  578022, 2021.

\bibitem[Alammary et~al.(2019{\natexlab{a}})Alammary, Alhazmi, Almasri, and
  Gillani]{b28}
A.~Alammary, S.~Alhazmi, M.~Almasri, and S.~Gillani, ``Blockchain-based
  applications in education: A systematic review,'' \emph{Applied Sciences},
  vol.~9, no.~12, p. 2400, 2019.

\bibitem[Loukil et~al.(2021)Loukil, Abed, and Boukadi]{b54}
F.~Loukil, M.~Abed, and K.~Boukadi, ``Blockchain adoption in education: A
  systematic literature review,'' \emph{Education and Information
  Technologies}, vol.~26, no.~5, pp. 5779--5797, 2021.

\bibitem[Mia et~al.(2022)Mia, Shahriar, Valero, Sakib, Saha, Barek, Faruk,
  Goodman, Khan, and Ahamed]{HIPPA}
M.~R. Mia, H.~Shahriar, M.~Valero, N.~Sakib, B.~Saha, M.~A. Barek, M.~J.~H.
  Faruk, B.~Goodman, R.~A. Khan, and S.~I. Ahamed, ``A comparative study on
  hipaa technical safeguards assessment of android mhealth applications,''
  \emph{Smart Health}, vol.~26, p. 100349, 2022.

\bibitem[Delgado-von Eitzen et~al.(2021{\natexlab{b}})Delgado-von Eitzen,
  Anido-Rif{\'o}n, and Fern{\'a}ndez-Iglesias]{b51}
C.~Delgado-von Eitzen, L.~Anido-Rif{\'o}n, and M.~J. Fern{\'a}ndez-Iglesias,
  ``Blockchain applications in education: A systematic literature review,''
  \emph{Applied Sciences}, vol.~11, no.~24, p. 11811, 2021.

\bibitem[Awaji et~al.(2020)Awaji, Solaiman, and Albshri]{b44}
B.~Awaji, E.~Solaiman, and A.~Albshri, ``Blockchain-based applications in
  higher education: A systematic mapping study,'' in \emph{Proceedings of the
  5th international conference on information and education innovations}, 2020,
  pp. 96--104.

\bibitem[Kishore et~al.(2021)Kishore, Chan, Muthupoltotage, Young, and
  Sundaram]{b71}
S.~Kishore, J.~Chan, U.~P. Muthupoltotage, N.~Young, and D.~Sundaram,
  ``Blockchain-based micro-credentials: Design, implementation, evaluation and
  adoption,'' in \emph{Hawaii International Conference on System
  Sciences}.\hskip 1em plus 0.5em minus 0.4em\relax Hawaii International
  Conference on System Sciences, 2021.

\bibitem[Yumna et~al.(2019{\natexlab{a}})Yumna, Khan, Ikram, and Ilyas]{b26}
H.~Yumna, M.~M. Khan, M.~Ikram, and S.~Ilyas, ``Use of blockchain in education:
  a systematic literature review,'' in \emph{Asian Conference on Intelligent
  Information and Database Systems}.\hskip 1em plus 0.5em minus 0.4em\relax
  Springer, 2019, pp. 191--202.

\bibitem[Saha(2023)]{saha2023analysis}
B.~Saha, ``Analysis of the adherence of mhealth applications to hipaa technical
  safeguards,'' 2023.

\bibitem[Turnbull et~al.(2020)Turnbull, Chugh, and Luck]{LCMS}
D.~Turnbull, R.~Chugh, and J.~Luck, ``Learning management systems, an
  overview,'' \emph{Encyclopedia of education and information technologies},
  pp. 1052--1058, 2020.

\bibitem[Ocheja et~al.(2019)Ocheja, Flanagan, Ueda, and Ogata]{LR}
P.~Ocheja, B.~Flanagan, H.~Ueda, and H.~Ogata, ``Managing lifelong learning
  records through blockchain,'' \emph{Research and Practice in Technology
  Enhanced Learning}, vol.~14, no.~1, pp. 1--19, 2019.

\bibitem[Apoki et~al.(2020)Apoki, Al-Chalabi, and Crisan]{meta}
U.~C. Apoki, H.~K.~M. Al-Chalabi, and G.~C. Crisan, ``From digital learning
  resources to adaptive learning objects: an overview,'' in \emph{Modelling and
  Development of Intelligent Systems: 6th International Conference, MDIS 2019,
  Sibiu, Romania, October 3--5, 2019, Revised Selected Papers}.\hskip 1em plus
  0.5em minus 0.4em\relax Springer, 2020, pp. 18--32.

\bibitem[Bhaskar et~al.(2021)Bhaskar, Tiwari, and Joshi]{b37}
P.~Bhaskar, C.~K. Tiwari, and A.~Joshi, ``Blockchain in education management:
  present and future applications,'' \emph{Interactive Technology and Smart
  Education}, vol.~18, no.~1, pp. 1--17, 2021.

\bibitem[Lombardi et~al.(2022)Lombardi, de~Villiers, Moscariello, and
  Pizzo]{auditing}
R.~Lombardi, C.~de~Villiers, N.~Moscariello, and M.~Pizzo, ``The disruption of
  blockchain in auditing--a systematic literature review and an agenda for
  future research,'' \emph{Accounting, Auditing \& Accountability Journal},
  vol.~35, no.~7, pp. 1534--1565, 2022.

\bibitem[Mohammad and Vargas(2022)]{usecase1}
A.~Mohammad and S.~Vargas, ``Barriers affecting higher education
  institutions’ adoption of blockchain technology: A qualitative study,'' in
  \emph{Informatics}, vol.~9, no.~3.\hskip 1em plus 0.5em minus 0.4em\relax
  MDPI, 2022, p.~64.

\bibitem[Shetty et~al.(2022)Shetty, Shetty, Pai, Rao, Bhandary, Shetty, Nayak,
  Keerthi~Dinesh, and Dsouza]{roi}
A.~Shetty, A.~D. Shetty, R.~Y. Pai, R.~R. Rao, R.~Bhandary, J.~Shetty,
  S.~Nayak, T.~Keerthi~Dinesh, and K.~J. Dsouza, ``Block chain application in
  insurance services: A systematic review of the evidence,'' \emph{SAGE Open},
  vol.~12, no.~1, p. 21582440221079877, 2022.

\bibitem[Mazzoni et~al.(2022)Mazzoni, Corradi, and Di~Nicola]{ConsenSys}
M.~Mazzoni, A.~Corradi, and V.~Di~Nicola, ``Performance evaluation of
  permissioned blockchains for financial applications: The consensys quorum
  case study,'' \emph{Blockchain: Research and applications}, vol.~3, no.~1, p.
  100026, 2022.

\bibitem[Huang et~al.(2020)Huang, Wang, Wu, Tyson, Luo, Zhang, Liu, Huang, and
  Jiang]{eosio}
Y.~Huang, H.~Wang, L.~Wu, G.~Tyson, X.~Luo, R.~Zhang, X.~Liu, G.~Huang, and
  X.~Jiang, ``Understanding (mis) behavior on the eosio blockchain,''
  \emph{Proceedings of the ACM on Measurement and Analysis of Computing
  Systems}, vol.~4, no.~2, pp. 1--28, 2020.

\bibitem[Tanana(2019)]{avalanche}
D.~Tanana, ``Avalanche blockchain protocol for distributed computing
  security,'' in \emph{2019 IEEE International Black Sea Conference on
  Communications and Networking (BlackSeaCom)}.\hskip 1em plus 0.5em minus
  0.4em\relax IEEE, 2019, pp. 1--3.

\bibitem[Br{\"u}njes and Gabbay(2020)]{Cardano}
L.~Br{\"u}njes and M.~J. Gabbay, ``Utxo-vs account-based smart contract
  blockchain programming paradigms,'' in \emph{Leveraging Applications of
  Formal Methods, Verification and Validation: Applications: 9th International
  Symposium on Leveraging Applications of Formal Methods, ISoLA 2020, Rhodes,
  Greece, October 20--30, 2020, Proceedings, Part III 9}.\hskip 1em plus 0.5em
  minus 0.4em\relax Springer, 2020, pp. 73--88.

\bibitem[Xu et~al.(2021)Xu, Sun, Luo, Cao, Yu, and
  Vasilakos]{hyperledger_fabric}
X.~Xu, G.~Sun, L.~Luo, H.~Cao, H.~Yu, and A.~V. Vasilakos, ``Latency
  performance modeling and analysis for hyperledger fabric blockchain
  network,'' \emph{Information Processing \& Management}, vol.~58, no.~1, p.
  102436, 2021.

\bibitem[Hajela et~al.(2021)Hajela, Pawar, and Phansalkar]{R3_Corda}
P.~Hajela, A.~Pawar, and S.~Phansalkar, ``Itreatu: An effective privacy and
  security solution for healthcare data using the r3 corda platform of
  blockchain technology,'' in \emph{Data Protection and Privacy in
  Healthcare}.\hskip 1em plus 0.5em minus 0.4em\relax CRC Press, 2021, pp.
  165--179.

\bibitem[Pierro and Tonelli(2022)]{solana}
G.~A. Pierro and R.~Tonelli, ``Can solana be the solution to the blockchain
  scalability problem?'' in \emph{2022 IEEE International Conference on
  Software Analysis, Evolution and Reengineering (SANER)}.\hskip 1em plus 0.5em
  minus 0.4em\relax IEEE, 2022, pp. 1219--1226.

\bibitem[Allouche et~al.(2021)Allouche, Frikha, Mitrea, Memmi, and
  Chaabane]{tezos}
M.~Allouche, T.~Frikha, M.~Mitrea, G.~Memmi, and F.~Chaabane, ``Lightweight
  blockchain processing. case study: scanned document tracking on tezos
  blockchain,'' \emph{Applied Sciences}, vol.~11, no.~15, p. 7169, 2021.

\bibitem[Widyastuti and Hermanto(2021)]{polka}
M.~Widyastuti and Y.~B. Hermanto, ``Cryptocurrency analysis of indonesian
  market education facilities,'' \emph{International Journal of Economics,
  Business and Accounting Research (IJEBAR)}, vol.~5, no.~2, 2021.

\bibitem[Priya et~al.(2020)Priya, Ponnavaikko, and Aantonny]{b68}
N.~Priya, M.~Ponnavaikko, and R.~Aantonny, ``An efficient system framework for
  managing identity in educational system based on blockchain technology,'' in
  \emph{2020 International Conference on Emerging Trends in Information
  Technology and Engineering (ic-ETITE)}.\hskip 1em plus 0.5em minus
  0.4em\relax IEEE, 2020, pp. 1--5.

\bibitem[Rahardja et~al.(2022)Rahardja, Aini, Khairunisa, Sunarya, and
  Millah]{b72}
U.~Rahardja, Q.~Aini, A.~Khairunisa, P.~A. Sunarya, and S.~Millah,
  ``Implementation of blockchain technology in learning management system
  (lms),'' \emph{APTISI Transactions on Management (ATM)}, vol.~6, no.~2, pp.
  112--120, 2022.

\bibitem[Xu et~al.(2023)Xu, Wang, and Jia]{b60}
J.~Xu, C.~Wang, and X.~Jia, ``A survey of blockchain consensus protocols,''
  \emph{ACM Computing Surveys}, 2023.

\bibitem[Rashid et~al.(2019)Rashid, Deo, Prasad, Singh, Chand, and Assaf]{b73}
M.~A. Rashid, K.~Deo, D.~Prasad, K.~Singh, S.~Chand, and M.~Assaf, ``Teduchain:
  A platform for crowdsourcing tertiary education fund using blockchain
  technology,'' \emph{arXiv preprint arXiv:1901.06327}, 2019.

\bibitem[Quigley(2022)]{80}
P.~Quigley, ``Blockchain primer,'' in \emph{Health Informatics}.\hskip 1em plus
  0.5em minus 0.4em\relax Productivity Press, 2022, pp. 297--330.

\bibitem[Palmisano et~al.(2022)Palmisano, Convertini, Sarcinella, Gabriele, and
  Bonifazi]{b75}
T.~Palmisano, V.~N. Convertini, L.~Sarcinella, L.~Gabriele, and M.~Bonifazi,
  ``Notarization and anti-plagiarism: A new blockchain approach,''
  \emph{Applied Sciences}, vol.~12, no.~1, p. 243, 2022.

\bibitem[Dias et~al.(2022)]{plagiarism}
P.~V. G.~F. Dias \emph{et~al.}, ``Blockchain and remote assessment system: a
  conceptual approach for plagiarism mitigation in the higher education
  institutions,'' Master's thesis, 2022.

\bibitem[Rojas et~al.(2021)Rojas, Gayoso~Mart{\'\i}nez, and Queiruga-Dios]{76}
W.~Rojas, V.~Gayoso~Mart{\'\i}nez, and A.~Queiruga-Dios, ``Blockchain in
  education: New challenges,'' in \emph{13th International Conference on
  Computational Intelligence in Security for Information Systems (CISIS 2020)
  12}.\hskip 1em plus 0.5em minus 0.4em\relax Springer, 2021, pp. 380--389.

\bibitem[Alam(2022)]{77}
A.~Alam, ``Platform utilising blockchain technology for elearning and online
  education for open sharing of academic proficiency and progress records,'' in
  \emph{Smart Data Intelligence: Proceedings of ICSMDI 2022}.\hskip 1em plus
  0.5em minus 0.4em\relax Springer, 2022, pp. 307--320.

\bibitem[Alam and Raza(2021)]{78}
M.~T. Alam and K.~Raza, ``Blockchain technology in healthcare: making digital
  healthcare reliable, more accurate, and revolutionary,'' in
  \emph{Translational Bioinformatics in Healthcare and Medicine}.\hskip 1em
  plus 0.5em minus 0.4em\relax Elsevier, 2021, pp. 81--96.

\bibitem[Lin et~al.(2022)Lin, Zhang, Li, Ji, and Sun]{79}
S.-Y. Lin, L.~Zhang, J.~Li, L.-l. Ji, and Y.~Sun, ``A survey of application
  research based on blockchain smart contract,'' \emph{Wireless Networks},
  vol.~28, no.~2, pp. 635--690, 2022.

\bibitem[Oyelere et~al.(2019)Oyelere, Tomczyk, Bouali, and Agbo]{b34}
S.~S. Oyelere, L.~Tomczyk, N.~Bouali, and F.~J. Agbo, ``Blockchain technology
  and gamification-conditions and opportunities for education,'' \emph{Adult
  Education 2018-Transformation in the Era of Digitization and Artificial
  Intelligence}, 2019.

\bibitem[Miyachi and Mackey(2021)]{81}
K.~Miyachi and T.~K. Mackey, ``hocbs: A privacy-preserving blockchain framework
  for healthcare data leveraging an on-chain and off-chain system design,''
  \emph{Information processing \& management}, vol.~58, no.~3, p. 102535, 2021.

\bibitem[Rahardja et~al.(2021)Rahardja, Aini, Oganda, Devana, et~al.]{82}
U.~Rahardja, Q.~Aini, F.~P. Oganda, V.~T. Devana \emph{et~al.}, ``Secure
  framework based on blockchain for e-learning during covid-19,'' in \emph{2021
  9th International Conference on Cyber and IT Service Management
  (CITSM)}.\hskip 1em plus 0.5em minus 0.4em\relax IEEE, 2021, pp. 1--7.

\bibitem[Alammary et~al.(2019{\natexlab{b}})Alammary, Alhazmi, Almasri, and
  Gillani]{b15}
A.~Alammary, S.~Alhazmi, M.~Almasri, and S.~Gillani, ``Blockchain-based
  applications in education: A systematic review,'' \emph{Applied Sciences},
  vol.~9, no.~12, p. 2400, 2019.

\bibitem[Zhang(2023)]{83}
M.~Zhang, ``Application of blockchain technology in colleges and universities
  financial sharing service platform,'' 2023.

\bibitem[Riad et~al.(2021)Riad, Islam, Shahriar, Zhang, Valero, Sneha, and
  Ahamed]{84}
A.~Riad, M.~S. Islam, H.~Shahriar, C.~Zhang, M.~Valero, S.~Sneha, and
  S.~Ahamed, ``Plugin-based tool for teaching secure mobile application
  development.'' \emph{Information Systems Education Journal}, vol.~19, no.~2,
  pp. 25--34, 2021.

\bibitem[Randolph et~al.(2022)Randolph, Faruk, Saha, Shahriar, Valero, Zhao,
  and Sakib]{randolph2022blockchain}
J.~Randolph, M.~J.~H. Faruk, B.~Saha, H.~Shahriar, M.~Valero, L.~Zhao, and
  N.~Sakib, ``Blockchain-based medical image sharing and automated
  critical-results notification: A novel framework,'' in \emph{2022 IEEE 46th
  Annual Computers, Software, and Applications Conference (COMPSAC)}.\hskip 1em
  plus 0.5em minus 0.4em\relax IEEE, 2022, pp. 1756--1761.

\bibitem[Fedorova and Skobleva(2020{\natexlab{b}})]{b39}
E.~P. Fedorova and E.~I. Skobleva, ``Application of blockchain technology in
  higher education.'' \emph{European Journal of Contemporary Education},
  vol.~9, no.~3, pp. 552--571, 2020.

\bibitem[Faruk et~al.(2022)Faruk, Saha, Islam, Alam, Shahriar, Valero, Rahman,
  Wu, and Alam]{faruk2022development}
M.~J.~H. Faruk, B.~Saha, M.~Islam, F.~Alam, H.~Shahriar, M.~Valero, A.~Rahman,
  F.~Wu, and Z.~Alam, ``Development of blockchain-based e-voting system:
  Requirements, design and security perspective,'' in \emph{2022 IEEE
  International Conference on Trust, Security and Privacy in Computing and
  Communications (TrustCom)}.\hskip 1em plus 0.5em minus 0.4em\relax IEEE,
  2022, pp. 959--967.

\bibitem[Alam and Benaida(2020)]{b42}
T.~Alam and M.~Benaida, ``Blockchain and internet of things in higher
  education,'' \emph{Tanweer Alam, Mohamed Benaida." Blockchain and Internet of
  Things in Higher Education." Universal Journal of Educational Research},
  vol.~8, pp. 2164--2174, 2020.

\bibitem[Haugsbakken et~al.(2019)Haugsbakken, Langseth, et~al.]{b33}
H.~Haugsbakken, I.~Langseth \emph{et~al.}, ``The blockchain challenge for
  higher education institutions,'' \emph{Eur. J. Educ}, vol.~2, no.~3, p.~41,
  2019.

\bibitem[Yumna et~al.(2019{\natexlab{b}})Yumna, Khan, Ikram, and Ilyas]{b3}
H.~Yumna, M.~M. Khan, M.~Ikram, and S.~Ilyas, ``Use of blockchain in education:
  a systematic literature review,'' in \emph{Asian Conference on Intelligent
  Information and Database Systems}.\hskip 1em plus 0.5em minus 0.4em\relax
  Springer, 2019, pp. 191--202.

\bibitem[Lizcano et~al.(2020)Lizcano, Lara, White, and Aljawarneh]{b31}
D.~Lizcano, J.~A. Lara, B.~White, and S.~Aljawarneh, ``Blockchain-based
  approach to create a model of trust in open and ubiquitous higher
  education,'' \emph{Journal of Computing in Higher Education}, vol.~32, no.~1,
  pp. 109--134, 2020.

\bibitem[Mittal et~al.(2021)Mittal, Gupta, Chaturvedi, Chansarkar, and
  Gupta]{peda}
A.~Mittal, M.~Gupta, M.~Chaturvedi, S.~R. Chansarkar, and S.~Gupta,
  ``Cybersecurity enhancement through blockchain training (cebt)--a serious
  game approach,'' \emph{International Journal of Information Management Data
  Insights}, vol.~1, no.~1, p. 100001, 2021.

\end{thebibliography}
\end{document}